\documentclass[12pt]{article}
\begin{document}
\title{EMERGENCE OF SPACETIME}
\author{B.G. Sidharth\\
International Institute for Applicable Mathematics \& Information Sciences\\
Hyderabad (India) \& Udine (Italy)\\
B.M. Birla Science Centre, Adarsh Nagar, Hyderabad - 500 063 (India)}
\date{}
\maketitle
\begin{abstract}
Starting from a background Zero Point Field (or Dark Energy) we show how an array of oscillators at the Planck scale leads to the formation of elementary particles and spacetime and also to a cosmology consistent with latest observations.
\end{abstract}
\section{Introduction}
Our starting point is the scenario in which there is particulate condensation in the all pervading Zero Point Field. This would be at the Compton length, which includes the Planck length as a special case. Indeed this was Einstein's belief. As Wilzeck
put it, ``Einstein was not satisfied with the dualism. He wanted to
regard the fields, or ethers, as primary. In his later work, he tried
to find a unified field theory, in which electrons (and of course
protons, and all other particles) would emerge as solutions in which
energy was especially concentrated, perhaps as singularities. But his
efforts in this direction did not lead to any tangible success.''\\
Let us see how this can happen. In the words of Wheeler,
`From the zero-point fluctuations of a single oscillator to the fluctuations 
of the electromagnetic field to geometrodynamic fluctuations is a natural 
order of progression...''\\ 
Let us consider,
following Wheeler a harmonic oscillator in its ground state. The
probability amplitude is
$$\psi (x) = \left(\frac{m\omega}{\pi \hbar}\right)^{1/4}
e^{-(m\omega/2\hbar)x^2}$$ for displacement by the distance $x$ from
its position of classical equilibrium. So the oscillator fluctuates
over an interval
$$\Delta x \sim (\hbar/m\omega)^{1/2}$$ The electromagnetic field is
an infinite collection of independent oscillators, with amplitudes
$X_1,X_2$ etc. The probability for the various oscillators to have
emplitudes $X_1, X_2$ and so on is the product of individual
oscillator amplitudes:
$$\psi (X_1,X_2,\cdots ) = exp [-(X^2_1 + X^2_2 + \cdots)]$$ wherein
there would be a suitable normalization factor. This expression gives
the probability amplitude $\psi$ for a configuration $B (x,y,z)$ of
the magnetic field that is described by the Fourier coefficients
$X_1,X_2,\cdots$ or directly in terms of the magnetic field
configuration itself by
$$\psi (B(x,y,z)) = P exp \left(-\int \int \frac{\bf{B}(x_1)\cdot
\bf{B}(x_2)}{16\pi^3\hbar cr^2_{12}} d^3x_1 d^3x_2\right).$$ $P$ being
a normalization factor. Let us consider a configuration where the
magnetic field is everywhere zero except in a region of dimension $l$,
where it is of the order of $\sim \Delta B$. The probability amplitude
for this configuration would be proportional to
$$\exp [-(\Delta B)^2 l^4/\hbar c)$$ So the energy
 of fluctuation in
a region of length $l$ is given by finally 
$$B^2 \sim \frac{\hbar c}{l^4}$$
 
In the above if $l$ is taken to be
the Compton wavelength of a typical elementary
 particle, then we
recover its energy $mc^2$, as can be easily verified.\\
In Quantum Gravity as well as in Quantum
Super String Theory, we encounter phenomena at a minimum scale. It is well known, and this was realized by Planck himself, that there is an absolute minimum scale in the universe, and this is,
$$l_P = \left(\frac{\hbar G}{c^3}\right)^{\frac{1}{2}} \sim
10^{-33}cm$$
\begin{equation}
t_P = \left(\frac{\hbar G}{c^5}\right)^{\frac{1}{2}} \sim
10^{-42}sec\label{ea1}
\end{equation}
Yet what we encounter in the real world is, not the Planck scale,
but the elementary particle Compton scale. The explanation given for
this is that the very high energy Planck scale is moderated by the
Uncertainty Principle. The question which arises is, exactly how
does this happen? We will now present an argument to show how the
Planck scale leads to the real world Compton scale, via
fluctuations and the modification of the Uncertainty Principle.\\
We note that (\ref{ea1}) defines the absolute minimum physical scale
\cite{bgsafdb,rr1,rr2,garay}. Associated with (\ref{ea1}) is the Planck mass
\begin{equation}
m_P \sim 10^{-5}gm\label{ea2}
\end{equation}
There are certain interesting properties associated with
(\ref{ea1}) and (\ref{ea2}). $l_P$ is the Schwarzschild radius of a
black hole of mass $m_P$ while $t_P$ is the evaporation time for
such a black hole via the Beckenstein radiation \cite{rr3}.
Interestingly $t_P$ is also the Compton time for the Planck mass,
a circumstance that is symptomatic of the fact that at this scale,
electromagnetism and gravitation become of the same order
\cite{cu}. Indeed all this fits in very well with Rosen's analysis
that such a Planck scale particle would be a mini universe
\cite{rr5,rr6}. We will now invoke a time varying gravitational
constant discussed extensively in ref.\cite{cu}.
\begin{equation}
G \approx \frac{lc^2}{m\sqrt{N}} \propto (\sqrt{N}t)^{-1} \propto
T^{-1}\label{ea3}
\end{equation}
which resembles the Dirac cosmology and features in the author's successful 1997 model, in which (\ref{ea3}) arises due to the fluctuation in the particle
number \cite{rr7,rr8,rr9,rr10,cu}. In (\ref{ea3}) $m$ and $l$ are the
mass and Compton wavelength of a typical elementary particle like
the pion while $N \sim 10^{80}$ is the number of elementary
particles in the universe, and
$T$ is the age of the universe.\\
In this scheme wherein (\ref{ea3}) follows from the theory, we use
the fact that given $N$ particles, the fluctuation in the particle
number is of the order $\sqrt{N}$, while a
typical time interval for the fluctuations is $\sim \hbar /mc^2$,
the Compton time. We will come back to this point later. So anticipating later work we have
$$\frac{dN}{dt} = \frac{\sqrt{N}}{\tau}$$
whence on integration we get,
$$T = \frac{\hbar}{mc^2} \sqrt{N}$$
and we can also deduce its spatial counterpart, $R = \sqrt{N} l$,
which is the well known empirical Eddington formula.\\
Equation (\ref{ea3}) which is an order of magnitude relation is
consistent with observation \cite{rr11,melnikov} while it may be remarked
that the Dirac cosmology itself has inconsistencies.\\
Substitution of (\ref{ea3}) in (\ref{ea1}) yields
$$l = N^{\frac{1}{4}} l_P,$$
\begin{equation}
t = N^{\frac{1}{4}} t_P\label{ea4}
\end{equation}
where $t$ as noted is the typical Compton time of an elementary
particle. We can easily verify that (\ref{ea4}) is correct. It
must be stressed that (\ref{ea4}) is not a fortuitous empirical
coincidence, but rather is a result of using (\ref{ea3}), which
again as noted, can be deduced from
theory.\\
(\ref{ea4}) can be rewritten as
$$l = \sqrt{n}l_P$$
\begin{equation}
t = \sqrt{n}t_P\label{ea5}
\end{equation}
wherein we have used (\ref{ea1}) and (\ref{ea3}) and $n = \sqrt{N}$.\\
We will now compare (\ref{ea5}) with the well known relations, referred to earlier,
\begin{equation}
R = \sqrt{N} l \quad T = \sqrt{N} t\label{ea6}
\end{equation}
The first relation of (\ref{ea6}) is the well known Weyl-Eddington
formula referred to while the second relation of (\ref{ea6}) is
given also on the right side of (\ref{ea3}). We now observe that
(\ref{ea6}) can be seen to be the result of a Brownian Walk
process, $l,t$ being typical intervals between "steps"
(Cf.\cite{cu,rr12,rr13}). We demonstrate this below after equation
(\ref{ea8}). On the other hand, the typical intervals $l,t$ can be
seen to result from a diffusion  process themselves. Let us
consider the well known diffusion relation,
\begin{equation}
(\Delta x)^2 \equiv l^2 = \frac{\hbar}{m} t \equiv \frac{\hbar}{m}
\Delta t\label{ea7}
\end{equation}
(Cf.\cite{rr12},\cite{rr14}-\cite{rr17}). What is being done here is that we are modeling fuzzy spacetime by a double Wiener process to be touched upon later, which leads to (\ref{ea7}). This will be seen in more detail, below.\\
Indeed as $l$ is the Compton wavelength, (\ref{ea7}) can be
rewritten as the Quantum Mechanical Uncertainty Principle
$$l \cdot p \sim \hbar$$
at the Compton scale (Cf. also \cite{rr18}) (or even at the de
Broglie
scale).\\
What (\ref{ea7}) shows is that a Brownian process defines
the Compton scale while (\ref{ea6}) shows that a Random Walk
process with the Compton scale as the interval defines the length
and time scales of the universe itself (Cf.\cite{rr13}). Returning
now to (\ref{ea5}), on using (\ref{ea2}), we observe that in
complete analogy with (\ref{ea7}) we have the relation
\begin{equation}
(\Delta x)^2 \equiv l^2_P = \frac{\hbar}{m_P} t_P \equiv
\frac{\hbar}{m_P} \Delta t\label{ea8}
\end{equation}
We can now argue that the Brownian process (\ref{ea8})
defines the Planck length while a Brownian Random Walk process
with the Planck scale as the interval leads to (\ref{ea5}), that is
the
Compton scale.\\
To see all this in greater detail, it may be observed that
equation (\ref{ea8}) (without subscripts)
\begin{equation}
(\Delta x)^2 = \frac{\hbar}{m} \Delta t\label{ea9}
\end{equation}
is the same as the equation (\ref{ea7}), indicative of a double
Wiener process. Indeed as noted by several scholars, this defines
the fractal Quantum path of
dimension 2 (rather than dimension 1) (Cf.e.g. ref.\cite{rr15}).\\
Firstly it must be pointed out that equation (\ref{ea9}) defines a
minimum space time unit - the Compton scale $(l,t)$. This follows
from (\ref{ea9}) if we substitute into it $\langle \frac{\Delta x}
{\Delta t}\rangle_{max} = c$. If the mass of the particle is the
Planck mass, then this Compton scale becomes the Planck scale.\\
Let us now consider the distance traversed by a particle with the
speed of light through the time interval $T$. The distance $R$
covered would be
\begin{equation}
\int dx = R = c \int dt = cT\label{eIa}
\end{equation}
by conventional reasoning. In view of the equation
(\ref{ea8}), however we would have to consider firstly, the minimum
time interval $t$ (Compton or Planck time), so that we have
\begin{equation}
\int dt \to nt\label{eIIa}
\end{equation}
Secondly, because the square of the space interval $\Delta x$
(rather than the interval $\Delta x$ itself as in conventional
theory) appears in (\ref{ea8}), the left side of (\ref{eIa})
becomes, on using (\ref{eIIa})
\begin{equation}
\int dx^2 \to \int (\sqrt{n}dx) (\sqrt{n}dy)\label{eIIIa}
\end{equation}
Whence for the linear dimension $R$ we would have
\begin{equation}
\sqrt{n}R = nct \quad \mbox{or} \quad R = \sqrt{n} l\label{eIVa}
\end{equation}
Equation (\ref{eIIIa}) brings out precisely the fractal dimension
$D = 2$ of the Brownian path while (\ref{eIVa}) is identical to
(\ref{ea4}) or (\ref{ea6}) (depending on whether we are dealing with
minimum intervals of the Planck scale or Compton scale of
elementary particles). Apart from showing the Brownian character
linking equations (\ref{ea4}) and (\ref{ea8}), incidentally, this
also provides the justification for what has so far been
considered to be a mysterious large number coincidence viz. the
Eddington
formula (\ref{ea6}).\\
There is another way of looking at this. It is well known that in approaches like that of the author or 
Quantum Super String Theory, at the Planck scale we have a non
commutative geometry encountered earlier \cite{rr19,bgsgrav}
Indeed this follows without recourse to Quantum
Super Strings, merely by the fact that $l_P,t_P$ are the absolute
minimum space
time intervals as shown by Snyder \cite{rr6}. as we saw earlier.\\
The non commutative geometry as is known, is
symptomatic of a modified uncertainty principle at this scale
\cite{rr22}-\cite{rr28}
\begin{equation}
\Delta x \approx \frac{\hbar}{\Delta p} + l^2_P \frac{\Delta
p}{\hbar}\label{ea10}
\end{equation}
The relation (\ref{ea10}) would be true even in Quantum Gravity.
The extra or second term on the right side of (\ref{ea10}) expresses the well known duality effect - as we attempt to go down
to the Planck scale, infact we are lead to the larger scale. The question is, what is this larger scale? If
we now use the fact that $\sqrt{n}$ is the fluctuation in the
number of Planck particles (exactly as $\sqrt{N}$ was the
fluctuation in the particle number as in (\ref{ea3})) so that
$\sqrt{n}mpc = \Delta p$ is the fluctuation or uncertainty in the
momentum for the second term on the right side of (\ref{ea10}), we
obtain for the uncertainty in length,
\begin{equation}
\Delta x = l^2_P \frac{\sqrt{n}m_Pc}{\hbar} =
l_P\sqrt{n},\label{ea11}
\end{equation}
We can easily see that (\ref{ea11}) is the same as the first
relation of (\ref{ea5}). The second relation of (\ref{ea5}) follows
from an
application of the time analogue of (\ref{ea10}).\\
Thus the impossibility of going down to the Planck scale because
of (\ref{ea10}), manifests itself in the fact that as
we attempt to go down to the Planck scale, we infact end up at the
Compton scale. This is how the Compton scale is encountered in real life.\\
Interestingly while at the Planck length, we have a life time of
the order of the Planck time, as noted above it is possible to
argue on the other hand that with the  mass and length of a
typical elementary particle like the pion, at the Compton scale,
we have a life time which is the age of the universe itself as
shown by
Sivaram \cite{rr3,rr29}.\\
Interestingly also Ng and Van Dam deduce the relations like
\cite{rr30}
\begin{equation}
\delta L \leq (Ll^2_P)^{1/3}, \delta T \leq
(Tt^2_P)^{1/3}\label{eax9}
\end{equation}
where the left side of (\ref{eax9}) represents the uncertainty in the measurement
of length and time for an interval $L,T$. We would like to point
out that if in the above we use for $L,T$, the size and age of
the universe, then $\Delta L$ and
$\Delta T$ reduce to the Compton scale $l,t$.\\
In conclusion, Brownian double Wiener processes and the modification
of the Uncertainity Principle at the Planck scale lead to the
physical Compton scale.
\section{The Universe as Planck Oscillators}
In the previous section, we had argued that a typical
elementary particle like a pion could be considered to be the result of
$n \sim 10^{40}$ evanescent Planck scale particles. We will return to this line of thinking again. The argument was based on 
random motions and also on the modification to the Uncertainity Principle.
We will now consider the problem from a different point of view,
which not only reconfirms the above result, but also enables an elegant
extension to the case of the entire universe itself.
Let us consider an array of $N$ particles, spaced a distance $\Delta x$
apart, which behave like oscillators, that is as if they were connected by
springs. We then have \cite{r2,r3}
\begin{equation}
r  = \sqrt{N \Delta x^2}\label{e1d}
\end{equation}
\begin{equation}
ka^2 \equiv k \Delta x^2 = \frac{1}{2}  k_B T\label{e2d}
\end{equation}
where $k_B$ is the Boltzmann constant, $T$ the temperature, $r$ the extent  and $k$ is the 
spring constant given by
\begin{equation}
\omega_0^2 = \frac{k}{m}\label{e3d}
\end{equation}
\begin{equation}
\omega = \left(\frac{k}{m}a^2\right)^{\frac{1}{2}} \frac{1}{r} = \omega_0
\frac{a}{r}\label{e4d}
\end{equation}
We now identify the particles with Planck masses, set $\Delta x \equiv a = 
l_P$, the Planck length. It may be immediately observed that use of 
(\ref{e3d}) and (\ref{e2d}) gives $k_B T \sim m_P c^2$, which ofcourse agrees 
with the temperature of a black hole of Planck mass. Indeed, as noted, Rosen had shown that a Planck mass particle at the Planck scale  can be considered to be a
universe in itself. We also use the fact alluded to that  a typical elementary particle
like the pion can be considered to be the result of $n \sim 10^{40}$ Planck
masses. Using this in (\ref{e1d}), we get $r \sim l$, the pion
Compton wavelength as required. Further, in this latter case, using (48) and the fact that $N = n \sim 10^{40}$, and (\ref{e2d}),i.e. $k_BT = kl^2/N$ and  (\ref{e3d}) and
(\ref{e4d}), we get for a pion, remembering that $m^2_P/n = m^2,$ 
$$k_ B T = \frac{m^3 c^4 l^2}{\hbar^2} = mc^2,$$
which of course is the well known formula for the Hagedorn temperature for
elementary particles like pions. In other words, this confirms the conclusions
in the previous section, that we can treat an elementary particle as a series of some
$10^{40}$ Planck mass oscillators. However it must be observed from 
(\ref{e2d}) and (\ref{e3d}), that while the Planck mass gives the highest
energy state, an elementary particle like the pion is in the lowest energy
state. This explains why we encounter elementary particles, rather than
Planck mass particles in nature. Infact as already noted \cite{cu}, a Planck
mass particle decays via the Bekenstein radiation within a Planck time
$\sim 10^{-42}secs$. On the other hand, the lifetime of an elementary particle
would be very much higher.\\
In any case the efficacy of our above oscillator model can be seen by the fact that we recover correctly the masses and Compton scales in the order of magnitude sense and also get the correct Bekenstein and Hagedorn formulas as seen above, and get the correct estimate of the mass of the universe itself, as will be seen below.\\
Using the fact that the universe consists of $N \sim 10^{80}$ elementary
particles like the pions, the question is, can we think of the universe as
a collection of $n N \, \mbox{or}\, 10^{120}$ Planck mass oscillators? This is what we will now
show. Infact if we use equation (\ref{e1d}) with
$$\bar N \sim 10^{120},$$
we can see that the extent $r \sim 10^{28}cms$ which is of the order of the diameter of the
universe itself. Next using (\ref{e4d}) we get
\begin{equation}
\hbar \omega_0^{(min)} \langle \frac{l_P}{10^{28}} \rangle^{-1} \approx m_P c^2 \times 10^{60} \approx Mc^2\label{e5d}
\end{equation}
which gives the correct mass $M$, of the universe which in contrast to the earlier pion case, is the highest energy state while the Planck oscillators individually are this time the lowest in this description. In other words the universe
itself can be considered to be described in terms of normal modes of Planck scale oscillators.\\
We will return to these considerations later: this and the preceeding considerations merely set the stage.\\
In the above cosmology of fluctuations, our starting
point was the creation
 of $\sqrt{N}$ particles within the minimum
time interval, a typical elementary
 particle Compton time $\tau$. A
rationale for this, very much in the spirit of
 the condensation of
particles from a background Zero Point Field as discussed can also be obtained in terms of a
 phase
transition from the Zero Point Field or Quantum Vacuum as we will see
in the sequel.
 In this case, particles are like the Benard cells
which form in fluids, as a result
 of a phase transition. While some
of the particles or cells may revert to
 the Zero Point Field, on the
whole there is a creation of $\sqrt{N}$ of these particles. If
 the
average time for the creation of the $\sqrt{N}$ particles or cells
is
 $\tau$, then at any point of time where there are $N$ such
particles, the time
 elapsed, in our case the age of the universe,
would be given by (\ref{ea6}) (Cf.
 \cite{r39,rr13}). While this is not
exactly the Big Bang scenario, there is
 nevertheless a rapid
creation of matter from the background Quantum Vacuum
 or Zero Point
Field. Thus over $10^{40}$ particles would have been
 created within
a fraction of a second.\\
 In any case when $\tau \to 0$, we recover
the Big Bang scenario with a singular
 creation of matter, while when
$\tau \to$ Planck time we recover the Prigogine
 Cosmology
(Cf.\cite{rr6} for details). However in neither of these two limits we
can
 deduce all the above consistent with observation large number
relations which therefore have to be branded as accidents.\\
 
The above cosmological model is related to the fact that there are
minimum
 space time intervals $l, \tau$. Indeed in this case as we
saw there is
 an underlying non commutative
geometry of spacetime \cite{r41,bgsgrav,r43} given by
\begin{equation}
[x,y] \approx 0(l^2),[x,p_x] = \imath \hbar [1 + \beta l^2], [t,E] =
\imath \hbar
 [1+ \gamma \tau^2]\label{e27}
\end{equation}
Interestingly (\ref{e27}) implies as we saw, modification to the usual
Uncertainty Principle. (This in turn has also been interpreted in
terms of a variable speed of light
 cosmology
\cite{r44,r45,r46}).\\
 
The relations (\ref{e27}), lead to the
modified Uncertainity relation
\begin{equation}
\Delta x \sim \frac{\hbar}{\Delta p} + \alpha' \frac{\Delta
p}{\hbar}\label{e28}
\end{equation}
(\ref{e28}) appears also in Quantum Super String Theory and is related
to the
 well known Duality relation
$$R \to \alpha'/ R$$
 (Cf.\cite{rr24,rr2}). In any case (\ref{e28}) is
symptomatic of the fact
 that we cannot go down to arbitrarily small
space time intervals. We now observe
 that the first term of
(\ref{e28}) gives the usual Uncertainity relation. In
 the second
term,
 we write $\Delta p = \Delta Nmc$,
 where $\Delta N$ is the
Uncertainity in the number of particles, in the
 universe. Also
$\Delta x = R$, the radius of the universe where
$$
 R \sim \sqrt{N}l,$$ 
the famous Eddington relationship (9). It
should be stressed that the otherwise
 emperical Eddington formula,
arises quite naturally in a Brownian characterisation
 of the
universe as has been pointed out in the previous Chapter (Cf. for
example ref.\cite{r2}). Put
 simply (9) is the Random Walk
equation.\\
  
We now get back,
$$\Delta N = \sqrt{N}$$
 This is the uncertainity in the particle
number, we used earlier.
 Substituting this in the time analogue of
the second term of (\ref{e28}), we
 immediately get, $T$ being the
age of the universe,
$$T = \sqrt{N} \tau$$ 
which is equation (\ref{ea6}). So, our
cosmology is self consistent with the
 modified relation
(\ref{e28}). The fluctuational effects are really couched in the
modification of the Heisenberg Principle, as given in (\ref{e28}).\\
Interestingly these minimum space time considerations can be related
to the
 Feynmann-Wheeler Instantaneous Action At a Distance
formulation (Cf.\cite{r50,r51,r52}), a point which we shall elaborate
further in the sequel.\\
 
We finally remark that relations like
(\ref{e27}) and (\ref{e28}), which can
 also be expressed in the
form, $a$ being the minimum length,
$$[x, p_x] = \imath \hbar [1 + \left(\frac{a}{\hbar}\right)^2 p^2]$$
(and can be considered to be truncated from a full series on the right
hand side
 (Cf. \cite{rr19}), could be deduced from the rather simple
model of a
 lattice - a one dimensional lattice for simplicity. In
this case we will have
 (Cf.\cite{rr6})
$$[x,p_x] = \imath \hbar cos \left(\frac{p}{\hbar} a\right),$$
 where
$a$ is the lattice length, $l$ the Compton length in our case. The
energy
 time relation now leads to a correction to the mass energy
formula, viz
$$E = mc^2 cos (kl), k \equiv p/\hbar$$
 
This is the contribution of
the extra term in the Uncertainity Principle.\\
 
As noted the Planck scale is an absolute minimum scale in
 the
universe. We argued that with the passage of time
the
 Planck scale would evolve to the present day elementary particle
Compton
 scale. To recapitulate: We have by definition
$$\hbar G/c^3 = l^2_P$$
 where $l_P$ is the Planck length $\sim
10^{-33}cms$. If we use $G$ from (\ref{ea3})
 in the above we will
get
\begin{equation}
l = N^{1/4} l_P\label{e29}
\end{equation}
Similarly we have
\begin{equation}
\tau = N^{1/4} \tau_P\label{e30}
\end{equation}
In (\ref{e29}) and (\ref{e30}) $l$ and $\tau$ denote the typical
elementary
 particle Compton length and time scale, and $N$ is the
number of such elementary
 particles in the universe.\\ 
We could explain these equations in terms of the
 Benard cell like elementary
particles referred to above. This time there are a
 total of $n =
\sqrt{N}$ Planck particles and (\ref{e29}) and (\ref{e30}) are
 the
analogues of equations (\ref{ea6}) in the context of
the
 formation of such particles. Indeed as we saw a Planck mass,
$m_P \sim 10^{-5}gms$, has a Compton life time and also a Bekenstein
Radiation
 life time of the order of the Planck time. These spacetime
scales are much
 too small and we encounter much too large energies
from the point of view of
 our experimental constraints. As noted in
the previous chapter our observed scale is the Compton scale,
 in
which Planck scale phenomena are moderated. In any case it can be seen
from
 the above that as the number of particles $N$ increases, the
scale evolves
 from the Planck to the Compton scale.\\
 
So, the
scenario which emerges is, that as the universe evolves, Planck
particles
 form the underpinning for elementary particles, which in
turn form the
 underpinning for the universe by being formed
continuously.\\
 
This can be confirmed by the following argument: We
can rewrite (\ref{e29})
 as
\begin{equation}
l = \nu' \sqrt{T}\label{e31}
\end{equation}
$$\nu' = l_P/\sqrt{\tau} \approx \hbar/m_P$$
 
Equation (\ref{e31}) as we saw earlier, is identical to
the
 Brownian diffusion process which is infact the underpinning for
equations like
 (\ref{e3}) or (\ref{e8}), except that this time we
have the same Brownian
 Theory operating from the Planck scale to the
Compton scale, instead of
 from the Compton scale to the edge of the
universe as seen above (Cf. also
 \cite{r2,ri}).\\
 
Interestingly,
let us apply the above scenario of $\sqrt{n}$ Planck particles
forming an elementary particle, to the extra term of the modified
Uncertainity
 Principle (\ref{e28}), as we did earlier in this
section in (iv). Remembering that
 $\alpha' = l^2_P$ in the theory,
and $\Delta p = N^{1/4} m_P c$, in this case,
 we get, as $\Delta x =
l$,
$$l = N^{1/4} l_P,$$
 which will be recognized as (\ref{e29}) itself!
Thus once again we see how the
 above cosmology is consistently tied
up with the non commutative space time
 expressed by equations
(\ref{e27}) or (\ref{e28}).\\
 
It may be mentioned that, as indeed
can be seen from (\ref{e29}) and (\ref{e30}),
 in this model, the
velocity of light remains constant.\\
 
We would now like to
comment further upon the Compton scale and the
 fluctuational
creation of particles alluded to above. In this case particles are
being
 produced out of a background Quantum Vacuum or Zero Point
Field
 which is pre space time. First a Brownian process
 alluded to
above defines the Planck length while a Brownian random
 process with
the Planck scale as the fundamental interval leads to
 the Compton
scale (Cf. also ref.\cite{garay}).\\
 
This process is a phase
transition, a critical phenomenon. To see
 this briefly, let us start
with the Landau-Ginsburg equation
 \cite{rr26}
\begin{equation}
- \frac{\hbar^2}{2m} \nabla^2 \psi + \beta |\psi |^2 \psi = -
 \alpha
\psi\label{ea7a}
\end{equation}
Here $\hbar$ and $m$ have the same meaning as in usual Quantum
Theory. It is remarkable that the above equation (\ref{ea7a}) is
identical with a similar Schrodinger like equation based on
amplitudes which is discussed in \cite{rr6}, where moreover $|\psi
|^2$ is proportional to the mass
 (or density) of the particle
(Cf. ref.\cite{rr6} for details). The
 equation in question is,
\begin{equation}
\imath \hbar \frac{\partial \psi}{\partial t} =
 \frac{-\hbar^2}{2m'}
\frac{\partial^2 \psi}{\partial x^2} + \int
 \psi^* (x')\psi (x) \psi
(x')U(x')dx',\label{ea8a}
\end{equation}
In (\ref{ea8a}), $\psi(x)$ is the probability of a particle being at
the point $x$ and the integral is over a region of the order of
 the
Compton wavelength. From this point of view, the similarity of
(\ref{ea8a}) with (\ref{ea7a}) need not be surprising considering also
that near critical points, due to universality diverse phenomena
like magnetism or fluids share similar mathematical equations.
Equation (\ref{ea8a}) was shown to lead to the Schrodinger equation
with the particle acquiring a mass (Cf.also ref.\cite{rr27}).\\
Infact in the Landau-Ginsburg case the coherence length is given
by
\begin{equation}
\xi = \left(\frac{\gamma}{\alpha}\right)^{\frac{1}{2}} = \frac{h
\nu_F} {\Delta}\label{ea9a}
\end{equation}
which can be easily shown to reduce to the Compton wavelength
(Cf. also ref.\cite{rr28}).\\
 Thus the emergence of Benard cell like
elementary particles from
 the Quantum Vacuum mimics the
Landau-Ginsburg phase transition. In
 this case we have a non local
growth of correlations reminiscent
 of the standard
 inflation
theory.\\
 As is known, the interesting aspects of the critical point
theory (Cf.ref.\cite{rr7}) are universality and scale. Broadly, this
means that diverse physical phenomena follow the same route at the
critical point, on the one hand, and on the other this can happen at
different scales, as exemplified for example, by the course graining
techniques of the Renormalization Group \cite{wilson}. To highlight
this point we note that in critical point phenomena we have the
reduced order parameter $\bar Q$ (which gives the fraction of the
excess of new states) and the reduced correlation length $\bar \xi$
(which follows from (\ref{ea9a})). Near the critical point we have
relations \cite{rr29} like
$$(\bar Q) = |t|^\beta , (\bar \xi) = |t|^{-\nu}$$ 
Whence
\begin{equation}
\bar Q^\nu = \bar \xi^\beta\label{e18a}
\end{equation}
In (\ref{e18a}) typically $\nu \approx 2\beta$. As $\bar Q \sim
\frac{1}{\sqrt{N}}$ because $\sqrt{N}$ particles are created
fluctuationally, given $N$ particles, and in view of the fractal two
dimensionality of the path
\begin{equation}
\bar Q \sim \frac{1}{\sqrt{N}}, \bar \xi = (l/R)^2\label{e19a}
\end{equation}
This gives the Eddington formula,
$$R = \sqrt{N}l$$ which is nothing but (\ref{ea6}).\\ There is another
way of looking at this. The noncommutative geometry (27) brings out
the primacy of the Quantum of Area. Indeed this has been noted from
the different perspective of Black Hole Thermodynamics too
\cite{baez}. We would also like to point out that a similar treatment
leads from the Planck scale to the Compton scale.\\ In other words the
creation of particles is the result of a critical point phase
transition and subsequent coarse graining (Cf. also
ref.\cite{baez}).\\ The above model apart from mimicking inflation
also explains as we saw, the
 so called miraculous large number
coincidences.\\ The peculiarity of these relations as we saw is that
they tie up large scale parameters like the radius or age of the
universe or the Hubble constant with microphysical parameters like the
mass, charge and the Compton scales of an elementary particle and the
gravitational constant. That is, the universe appears to have a
Machian or holistic feature. One way to understand why the large and
the small are tied up is to remember, as we saw in the earlier
chapter, that there is an underpinning of normal mode Planck
oscillators, that is, collective phenomena all across the universe.\\
\section{The Nature of Space Time}
As we noted earlier all of Classical Physics and Quantum Theory, is
based on the Minkowski spacetime, as for example in the case of
Quantum Field Theory, or Reimannian spacetime as in the case of
General Relativity. In the non relativistic theories, Newtonian
spacetime, is used, which is a special case of Minkowskian
spacetime. But in all these cases the common denominator is that we
are dealing with a differentiable manifold.\\ 
This breaks down however in Quantum Gravity
including the author's approach, String Theory and other approaches,
be it at the Planck scale, or at the Compton scale
\cite{rr22,rr1,bib3,bib4}. The underlying reason for this breakdown
of a differentiable spacetime manifold is the Uncertainty
Principle--as we go down to arbitrarily small spacetime intervals, we
encounter arbitrarily large energy momenta. As Wheeler put it
\cite{rr6}, ``no prediction of spacetime, therefore no meaning for
spacetime is the verdict of the Quantum Principle. That object which
is central to all of Classical General Relativity, the four
dimensional spacetime geometry, simply does not exist, except in a
classical approximation.'' Before proceeding to analyse the nature of
spacetime beyond the classical approximation, let us first analyse
briefly the nature of classical spacetime itself.\\ We can get an
insight into the nature of the usual spacetime by considering the well
known formulation of Quantum Theory in terms of stochastic processes
more precisely, a double Wiener process which, as we saw, models fuzzy
spacetime \cite{rr14,rr18}.\\ In the stochastic approach, we
deal with a double Wiener process which leads to a complex velocity
$V-\imath U$. It is this complex velocity that leads to Quantum Theory
from the usual diffusion theory as seen in the previous Chapter.\\ To
see this in a simple way, let us write the usual diffusion equation as
\begin{equation}
\Delta x \cdot \Delta x = \frac{h}{m}\Delta t \equiv \nu \Delta
t\label{eb1}
\end{equation}
We saw that equation (\ref{eb1}) can be rewritten as the usual Quantum
Mechanical relation,
\begin{equation}
m\frac{\Delta x}{\Delta t} \cdot \Delta x = h = \Delta p \cdot \Delta
x\label{eb2}
\end{equation}
We are dealing here, with phenomena within the Compton or de Broglie
wavelength.\\ We now treat the diffusion constant $\nu$ to be very
small, but non vanishing. That is, we consider the semi classical
case. This is because, a purely classical description, does not
provide any insight.\\ It is well known that in this situation we can
use the WKB approximation \cite{bib7}. Whence the right hand side of
the wave function,
$$\psi = \sqrt{\rho} e^{\imath /\hbar S}$$ goes over to, in the one
dimensional case, for simplicity,
$$(p_x) ^{-\frac{1}{2}} e^{\frac{1}{h}} \int p(x)dx$$ so that we have,
on comparison,
\begin{equation}
\rho = \frac{1}{p_x}\label{eb3}
\end{equation}
$\rho$ being the probability density. In this case the condition $U
\approx 0$, that is, the velocity potential becoming real, implies
\begin{equation}
\nu \cdot \nabla ln (\sqrt{\rho}) \approx 0\label{eb4}
\end{equation}
This semi classical analysis suggests that $\sqrt{\rho}$ is a slowly
varying function of $x$, infact each of the factors on the left side
of (\ref{eb4}) would be $\sim 0(h)$, so that the left side is $\sim
0(h^2)$ (which is being neglected).  Then from (\ref{eb3}) we conclude
that $p_x$ is independent of $x$, or is a slowly varying function of
$x$. The equation of continuity now gives
$$\frac{\partial \rho}{\partial t} + \vec \nabla (\rho \vec v) =
\frac{\partial \rho} {\partial t} = 0$$ That is the probability
density $\rho$ is independent or nearly so, not only of $x$, but also
of $t$. We are thus in a stationary and homogenous scenario. This is
strictly speaking, possible only in a single particle universe, or for
a completely isolated particle, without any effect of the
environment. Under these circumstances we have the various
conservation laws and the time reversible theory, all this taken over
into Quantum Mechanics as well. This is an approximation valid for
small, incremental changes, as indeed is implicit in the concept of a
differentiable space time manifold.\\ Infact the well known
displacement operators of Quantum Theory which define the energy
momentum operators are legitimate and further the energy and momenta
are on the same footing only under this approximation\cite{bib8}.\\ We
would now like to point out the well known close similarity between
the formulation mentioned above and the hydrodynamical formulation for
Quantum Mechanics, which also leads to identical equations on writing
the wave function as above. These two approaches were reconciled by
considering quantized vortices at the Compton scale
(Cf.\cite{rr12}).\\ To proceed further, we start with the
Schrodinger equation
\begin{equation}
\imath \hbar \frac{\partial \psi}{\partial t} = - \frac{\hbar^2}{2m}
\nabla^2 \psi + V \psi\label{eb5}
\end{equation}
Remembering that for momentum eigen states we have, for simplicity,
for one dimension
\begin{equation}
\frac{\hbar}{\imath} \frac{\partial}{\partial x} \psi =
p\psi\label{eb6}
\end{equation}
where $p$ is the momentum or $p/m$ is the velocity $v$, we take the
derivative with respect to $x$ (or $\vec x$) of both sides of
(\ref{eb5}) to obtain, on using (\ref{eb6}),
\begin{equation}
\imath \hbar \frac{\partial (v\psi )}{\partial t} = -
\frac{\hbar^2}{2m} \nabla^2 (v \psi) + \frac{\partial V}{\partial x}
\psi + Vv\psi\label{eb7}
\end{equation}
We would like to compare (\ref{eb7}) with the well known equation for
the velocity in hydrodynamics\cite{bib10,bib11}, following from the
Navier-Stokes equation,
\begin{equation}
\rho \frac{\partial v}{\partial t} = -\nabla p - \rho \alpha T g + \mu
\nabla^2 v\label{eb8}
\end{equation}
In (\ref{eb8}) $v$ is a small perturbational velocity in otherwise
stationary flow between parallel plates separated by a distance $d, p$
is a small pressure, $\rho$ is the density of the fluid, $T$ is the
temperature proportional to $Q(z)v,\mu$ is the Navier-Stokes
coefficient and $\alpha$ is the coefficient of volume expansion with
temperature. Also required would be
$$\beta \equiv \frac{\Delta T}{d}.$$ $v$ itself is given by
\begin{equation}
v_z = W(z)exp (\sigma t + \imath k_xx + \imath k_y y),\label{eb9}
\end{equation}
$z$ being the coordinate perpendicular to the fluid flow.\\ We can now
see the parallel between equations (\ref{eb7}) and (\ref{eb8}). To
verify the identification we would require that the dimensionless
Rayleigh number
$$R = \frac{\alpha \beta g d^4}{\kappa \nu}$$ should have an analogue
in (\ref{eb7}) which is dimensionless, $\kappa , \nu$ being the
thermometric conductivity and viscocity.\\ Remembering that
$$\nu \sim \frac{h}{m}$$ and
$$d \sim \lambda$$ where $\lambda$ is the Compton wavelength in the
above theory (Cf.\cite{bib12} for details) and further we have
\begin{equation}
\rho \propto f(z)g = V\label{eb10}
\end{equation}
for the identification between the hydrostatic energy and the energy
$V$ of Quantum Mechanics, it is easy using (\ref{eb10}) and earlier
relations to show that the analogue of $R$ is
\begin{equation}
(c^2/\lambda^2) \cdot \lambda^4 \cdot (m/h)^2\label{eb11}
\end{equation}
The expression (\ref{eb11}) indeed is dimensionless and of order
$1$. Thus the mathematical identification is complete.\\ 
Before
proceeding, let us look at the physical significance of the above
considerations (Cf.\cite{bib13} for a graphic description.) Under
conditions of stationery flow, when the diifference in the temperature
between the two plates is negligible there is total translational
symmetry, as in the case of the displacement operators of Quantum
Theory. But when there is a small perturbation in the velocity (or
equivalently the temperature difference), then beyond a critical value
the stationarity and homogeneity of the fluid is disrupted, or the
symmetry is broken and we have the phenomena of the formation of
Benard cells, which are convective vortices and can be counted. This
infact is the "birth" of space It must be stressed that before the
formation of the Benard cells, there is no ``space'', that is, no
point to distinguish from or relate to another point. Only with the
formation of the cells are we able to label space points.\\ 
In the
context of the above identification, the Benard cells would correspond
to the formation of ``quantized vortices'' at the Compton (Planck)
scale from the ZPF, as we saw, which latter had been discussed in
detail in the literature (Cf.\cite{bib14}) from the
ZPF. This phase transition would correspond to the ``formation'' of
spacetime. As discussed in detail these ``quantized
vortices'' can be identified with elementary particles. Interestingly,
as noted Einstein himself considered electrons as condensates from a
background electromagnetic field. All this ties up with the
discussion in the previous section.\\ 
However in order to demonstrate
that the above formulation is not a mere mathematical analogy, we have
to show that the critical value of the wave number $k$ in the
expression for the velocity in the hydrodynamical flow (\ref{eb9}) is
the same as the critical length, the Compton length. In terms of the
dimensionless wave number $k' = k/d$, this critical value is given
by\cite{bib10}
$$k'_c \sim 1$$ In the case of the ``quantized vortices'' at the
Compton scale $l$, remembering that $d$ is identified with $l$ itself
we have,
$$l'_c (\equiv) k'_c \sim 1,$$ exactly as required.\\ 
In this
connection it may be mentioned that due to fluctuations in the Zero
Point Field or the Quantum Vaccuum, there would be fluctuations in the
metric given by as is known,
$$\Delta g \sim l_P/l$$ 
where $l_P$ is the Planck length and $l$ is a
small interval under consideration. At the same time the fluctuation
in the curvature of space would be given by
$$\Delta R \sim l_P/l^3$$ Normally these fluctuations are extremely
small but as discussed in detail elsewhere\cite{r7}, this would imply
that at the Compton scale of a typical elementary particle $l \sim
10^{-11}cms$, the fluctuation in the curvature would be $\sim 1$. This
is symptomatic of the formation of what we have termed above as
elementary particle ``quantized vortices''.\\ Further if a typical
time interval between the formation of such ``quantized vortices''
which are the analogues of the Benard cells is $\tau$, in this case
the Compton time, then as in the theory of the Brownian Random
Walk\cite{rief}, the mean time extent would be given by
\begin{equation}
T \sim \sqrt{N}\tau\label{eb12}
\end{equation}
where $N$ is the number of such quantized vortices or elementary
particles (Cf.also \cite{rr12}). This is equation (\ref{ea6}) - that is,
the equation (\ref{eb12}) holds good for the universe itself because
$T$ the age of the universe $\sim 10^{17} secs$ and $N$ the number of
elementary particles $\sim 10^{80}, \tau$ being the Compton time $\sim
10^{-23} secs$. Interestingly, this ``phase transition'' nature of
time would automatically make it irreversible, unlike the conventional
model in which time is reversible. We will return to these
considerations later in this section.\\ It may be mentioned that an
equation similar to (\ref{eb12}) can be deduced by the same arguments
for space extension also as indeed we did, and this time we get back
the well known Eddington formula viz.,
\begin{equation}
R \sim \sqrt{N} l\label{eb13}
\end{equation}
where $R$ is the extent or radius of the universe and $l$ is the cell
size or Compton wavelength. We can similarly characterize the
formation of elementary particles themselves from cells at the Planck
scale.\\ Once we recognize the minimum space time extensions, then we
immediately are lead to the underlying non commutative geometry
encountered in the earlier chapter and given by equation (27):
\begin{equation}
[x,y] = 0(l^2),[x,p_x] = \imath \hbar [1+0(l^2)], [t,E]=\imath \hbar
[1 + 0(\tau^2)\label{eb14}
\end{equation}
As we saw relations like (27) are Lorentz invariant.  At this stage we
recognise the nature of spacetime as given by (27) in contrast to the
stationary and homogeneous spacetime discussed earlier.
Witten\cite{rr24,bib16} has called this Fermionic spacetime as
contrasted with the usual Bosonic spacetime. Indeed we traced the
origins of the Dirac equation of the electron to (27). We also argued
that (27) provides the long sought after reconciliation between
electromagnetism and gravitation\cite{bgsgrav,r52}.\\ The usual
differentiable spacetime geometry can be obtained from (27) if $l^2$
is neglected, and this is the approximation that has been implicit.\\
Thus spacetime is a collection of such cells or elementary particles.
As pointed out earlier, this spacetime emerges from a homogeneous
stationary non or pre spacetime when the symmetry is broken, through
random processes. The question that comes up then is, what is the
metric which we use? This has been touched upon earlier, and we will
examine it again.\\ 
We first make a few preliminary remarks. When we
talk of a metric or the distance between two "points" or "particles",
a concept that is implicit is that of topological "nearness" - we
require an underpinning of a suitably large number of "open"
sets\cite{bib18}. Let us now abandon the absolute or background space
time and consider, for simplicity, a universe (or set) that consists
solely of two particles. The question of the distance between these
particles (quite apart from the question of the observer) becomes
meaningless. Indeed, this is so for a universe consisting of a finite
number of particles. For, we could isolate any two of them, and the
distance between them would have no meaning.  We can intuitiively
appreciate that we would infact need distances of intermediate or more
generally, other points.\\ In earlier work\cite{bib19,bib20},
motivated by physical considerations we had considered a series of
nested sets or neighbourhoods which were countable and also whose
union was a complete Hausdorff space. The Urysohn Theorem was then
invoked and it was shown that the space of the subsets was metrizable.
Let us examine this in more detail.\\ Firstly we observe that in the
light of the above remarks, the concepts of open sets, connectedness
and the like reenter in which case such an isolation of two points
would not be possible.\\ More formally let us define a neighbourhood
of a particle (or point or element) $A$ of a set of particles as a
subset which contains $A$ and atleast one other distinct element. Now,
given two particles (or points) or sets of points $A$ and $B$, let us
consider a neighbourhood containing both of them, $n(A,B)$ say. We
require a non empty set containing atleast one of $A$ and $B$ and
atleast one other particle $C$, such that $n(A,B) \subset n(A,C)$, and
so on. Strictly, this "nested" sequence should not terminate. For, if
it does, then we end up with a set $n(A,P)$ consisting of two isolated
"particles" or points, and the "distance" $d(A,P)$ is meaningless.\\
We now assume the following property\cite{bib19,bib20}: Given two
distinct elements (or even subsets) $A$ and $B$, there is a
neighbourhood $N_{A_1}$ such that $A$ belongs to $N_{A_1}$, $B$ does
not belong to $N_{A_1}$ and also given any $N_{A_1}$, there exists a
neighbourhood $N_{A_\frac{1}{2}}$ such that $A \subset
N_{A_\frac{1}{2}} \subset N_{A_1}$, that is there exists an infinite
topological closeness.\\ From here, as in the derivation of Urysohn's
lemma\cite{bib21}, we could define a mapping $f$ such that $f(A) = 0$
and $f(B) = 1$ and which takes on all intermediate values. We could
now define a metric, $d(A,B) = |f(A) - f(B)|$. We could easily verify
that this satisfies the properties of a metric.\\ With the same
motivation we will now deduce a similar result, but with different
conditions. In the sequel, by a subset we will mean a proper subset,
which is also non null, unless specifically mentioned to be so. We
will also consider Borel sets, that is the set itself (and its
subsets) has a countable covering with subsets. We then follow a
pattern similar to that of a Cantor ternary set \cite{bib18,bib22}. So
starting with the set $N$ we consider a subset $N_1$ which is one of
the members of the covering of $N$ and iterate this process so that
$N_{12}$ denotes a subset belonging to the covering of $N_1$ and so
on.\\ We note that each element of $N$ would be contained in one of
the series of subsets of a sub cover. For, if we consider the case
where the element $p$ belongs to some $N_{12\cdots m}$ but not to
$N_{1,2,3\cdots m+1}$, this would be impossible because the latter
form a cover of the former. In any case as in the derivation of the
Cantor set, we can put the above countable series of sub sets of sub
covers in a one to one correspondence with suitable sub intervals of a
real interval $(a,b)$.\\ {\large \bf{Case I}}\\ If $N_{1,2,3\cdots m}
\to$ an element of the set $N$ as $m \to \infty$, that is if the set
is closed, we would be establishing a one to one relationship with
points on the interval $(a,b)$ and hence could use the metric of this
latter interval, as seen earlier.\\ {\large \bf{Case II}}\\ It is
interesting to consider the case where in the above iterative
countable process, the limit does not tend to an element of the set
$N$, that is set $N$ is not closed and has what we may call singular
points. We could still truncate the process at $N_{1,2,3\cdots m}$ for
some $m > L$ arbitrary and establish a one to one relationship between
such truncated subsets and arbitrarily small intervals in $a,b$. We
could still speak of a metric or distance between two such
arbitrarily small intervals.\\ This case is of interest because we
described elementary particles as, what we have called Quantum
Mechanical Kerr-Newman Black Holes or vortices, where we have a length
of the order of the Compton wavelength as seen in the previous
sections, within which spacetime as we know it breaks down. Such cut
offs as seen lead to a non commutative geometry (27) and what may be
called fuzzy spaces\cite{bib23,rr1}.(We note that the centre of the
vortex is a singular point). In any case, the number of particles in
the universe is of the order $10^{80}$, which approxiimates infinity
from a physicist's point of view.\\ Interestingly, we usually consider
two types of infinite sets - those with cardinal number $n$
corresponding to countable infinities, and those with cardinal number
$c$ corresponding to a continuum, there being nothing in between
\cite{bib22}. This is the well known but unproven Continuum
Hypotheses.\\ What we have shown with the above process is that it is
possible to concieve an intermediate possibility with a cardinal
number $n^p, p > 1$.\\ In the above considerations three properties
are important: the set must be closed i.e. it must contain all its
limit points, perfect i.e. in addition each of its points must be a
limit point and disconnected i.e. it contains no nonnull open
intervals. Only the first was invoked in Case I.\\ Finally we notice
again the holistic feature. A metric emerges by considering large
encompassing sets. It may be remarked that much of Quantum Theory,
like much of Classical Theory was couched in the concepts of Newtonian
two body mechanics and determinism. The moment we consider even a
three body problem, as was realized by Poincare more than a century
ago, the picture gets altered. As he noted \cite{bib27}, ``If we knew
exactly the laws of nature and the situation of the universe at the
initial moment, we could predict exactly the situation of that same
universe at a succeeding moment. But even if it were the case that the
natural laws had no longer any secret for us, we could still know the
situation approximately. If that enabled us to predict the succeeding
situation with the same approximation, that is all we require, and we
should say that the phenomenon had been predicted, that it is governed
by the laws. But it is not always so; it may happen that small
differences in the initial conditions produce very great ones in the
final phenomena. A small error in the former will produce an enormous
error in the latter. Prediction becomes impossible.''  In a similar
vein, Prigogine observes \cite{bib13}, ``Our physical world is no
longer symbolized by the stable and periodic planetary motions that
are at the heart of classical mechanics. It is a world of
instabilities and fluctuations...''\\ Indeed, the departure from the
two body formulation began with electromagnetism itself, which has to
invoke the environment.\\ We now return to the current view of Planck
scale oscillators in the background dark energy or Quantum Vaccuum. In
this context we saw that elementary particles can
be considered to be normal modes of $n \sim 10^{40}$ Planck
oscillators in the ground state, while the etire universe itself has
an underpinning of $\bar {N} \sim 10^{120}$ such oscillators, there
being $N \sim 10^{80}$ elementary particles in the universe
\cite{r2,rr25}. These Planck oscillators are formed out of the Quantum
Vaccuum (or dark energy). Thus we have, $m_P c^2$ being the energy of
each Planck oscillator, $m_P$ being the Planck mass, $\sim
10^{-5}gms$,
\begin{equation}
m = \frac{m_P}{\sqrt{n}}\label{ey}
\end{equation}
\begin{equation}
l = \sqrt{n} l_P, \tau = \sqrt{n} \tau_P , n = \sqrt{N}\label{ex1}
\end{equation}
where $m$ is the mass of a typical elementary particle, taken to be a
pion in the literature. The ground state of $\bar {N}$ such Planck
oscillators would be, in analogy to (\ref{ey}),
\begin{equation}
\bar {m} = \frac{m_P}{\sqrt{N}} \sim 10^{-65}gms\label{ex2}
\end{equation}
The universe is an excited state and consists of $N$ such ground state
levels and so we have, from (\ref{ex2})
$$M = \bar{m} N = \sqrt{N} m_P \sim 10^{55}gms,$$ as required, $M$
being the mass of the universe.\\ 
Due to the fluctuation $\sim
\sqrt{n}$ in the levels of the $n$ oscillators making up an elementary
particle, the energy is, remembering that $mc^2$ is the ground state,
$$\Delta E \sim \sqrt{n} mc^2 = m_P c^2,$$ using (\ref{ex1}), and so
the indeterminacy time is
$$\frac{\hbar}{\Delta E} = \frac{\hbar}{m_Pc^2} = \tau_P,$$ as indeed
we would expect.\\ The corresponding minimum indeterminacy length
would therefore be $l_P$.  One of the consequences of the minimum
spacetime cut off as we saw is that the Heisenberg Uncertainty
Principle takes an extra term as we saw previously \cite{mup}. Thus as we saw
\begin{equation}
\Delta x \approx \frac{\hbar}{\Delta p} + \alpha \frac{\Delta
p}{\hbar},\, \alpha = l^2 (\mbox{or}\, l^2_P)\label{ex6}
\end{equation}
where $l$ (or $l_P$) is the minimum interval under consideration. This
is just (28).  The first term gives the usual Heisenberg Uncertainty
Principle.\\ Application of the time analogue of (\ref{ex6}) for the
indeterminacy time $\Delta t$ for the fluctuation in energy $\Delta
\bar{E} = \sqrt{N} mc^2$ in the $N$ particle states of the universe
gives exactly as above,
$$\Delta t = \frac{\Delta E}{\hbar} \tau^2_P =
\frac{\sqrt{N}mc^2}{\hbar} \tau^2_P = \frac{\sqrt{N}
m_Pc^2}{\sqrt{n}\hbar} \tau^2_P = \sqrt{n} \tau_P = \tau ,$$ wherein
we have used (\ref{ex1}). In other words, for the fluctuation
$\sqrt{N}$, the time is $\tau$. It must be re-emphasized that the
Compton time $\tau$ of an elementary particle, is an interval within
which there are unphysical effects like zitterbewegung - as pointed
out by Dirac, it is only on averaging over this interval, that we
return to meaningful Physics.  This gives us,
\begin{equation}
dN/dt = \sqrt{N}/\tau\label{ex3}
\end{equation}
Equation (78) is identical to (3), the starting point for the
cosmology discussed. Here we have derived this relation from a
consideration of the underlying Planck oscillators.  On the other hand
due to the fluctuation in the $\sqrt {\bar{N}}$ oscillators
constituting the universe, the fluctuational energy is similarly given
by
$$\sqrt{\bar {N}} \bar {m} c^2,$$ which is the same as (\ref{ex2})
above. Another way of deriving (\ref{ex3}) is to observe that as
$\sqrt{n}$ particles appear fluctuationally in time $\tau_P$ which is,
in the elementary particle time scales, $\sqrt{n} \sqrt{n} = \sqrt{N}$
particles in $\sqrt{n} \tau_P = \tau$. That is, the rate of the
fluctuational appearance of particles is
$$
\left(\frac{\sqrt{n}}{\tau_P}\right) = \frac{\sqrt{N}}{\tau} = dN/dt$$
which is (\ref{ex3}). From here by integration,
$$T = \sqrt{N} \tau$$ $T$ is the time elapsed from $N = 1$ and $\tau$
is the Compton time. This gives $T$ its origin in the fluctuations -
there is no smooth ``background'' (or ``being'') time - the root of
time is in ``becoming''. It is the time of a Brownian double Wiener
process: A step $l$ gives a step in time $l/c \equiv \tau$ and
therefore $\Delta x = \sqrt{N} l$ gives $T = \sqrt{N} \tau$. Time is
born out of acausal fluctuations which are random and therefore
irreversible. Indeed, there is no background time. Time is
proportional to $\sqrt{N}$, $N$ being the number of particles which
are being created spontaneously from the ZPF.\\ The time we use is
what may be called Stationary time and it is an approximation as we
saw \cite{rr18}. Further, Quantum Mechanics, Gravitation etc. follow
from here. In Quantum Mechanics, the measurement of the observer
triggers the acausal collapse of the wave function - an irreversible
event - but the wave function itself satisfies a deterministic and
reversible equation paradoxically. Yet the universe is
``irreversible''. It appears spontaneous irreversibility or
indeterministic time \cite{rr12} is the real time. This can be
contrasted to the usual time reversible Quantum Theory. \\ We observe
that \cite{land}
$$\psi (r,t) = \frac{1}{(2\pi \hbar)^{3/2}} \int a(p) \exp
\left[\frac{\imath}{\hbar} \left(p \cdot r - \frac{p^2}{2m}
t\right)\right] dp,$$ $a(p)$ being independent of time. So we have at
any other time $t'$:
$$a(p) = \frac{1}{(2\pi \hbar)^{3/2}} \int \psi (r',t') \exp
\left[-\frac{\imath}{\hbar}\left(p \cdot r' - \frac{p^2}{2m}
t'\right)\right] dr'$$ Substitution yields the result
\begin{equation}
\psi (r,t) = \int K(r,t'; r',t') \psi (r',t') dr',\label{exx}
\end{equation}
the Kernel function $K$ being given by
$$K(r,t; r',t') = \frac{1}{(2\pi \hbar)^3} \int \exp
\left\{\frac{\imath}{\hbar}\left[ p \cdot (r-r') - \frac{p^2}{2m} (t -
t')\right]\right\} dp$$ or after some manipulation, in the form
$$K (r,t; r',t') = \left[ \frac{2\pi \imath \hbar}{m} (t - t')
\right]^{-3/2} \exp \left[\imath \frac{m}{2\hbar} \frac{|r - r'|^2}{(t
- t')}\right]$$ The point is that in (\ref{exx}) $\psi (r,t)$ at $t$
is given in terms of a linear expansion of $\psi (r,t')$ at earlier
times $t'$. But what is to be noted is, the symmetry between $t$ and
$t'$. This is not surprising as the original Schrodinger equation
remains unchanged under $t \to -t$.\\ Thus it is possible to
understand the fluctuations encountered in Section 2, that is, the
equation (78) which was the starting point for fluctuational energy in
terms of the underpinning of Planck scale oscillators in the Quantum
Vaccuum.\\ We would now like to make some remarks. Starting from a
completely different point of view namely Black Hole Thermodynamics,
Landsberg \cite{pc} deduced that the smallest observable mass in the
universe is $\sim 10^{-65}gms$, which is exactly the minimum mass
given in (\ref{ex2}).\\ Further due to the fluctuational appearance of
$\sqrt{N}$ particles, the fluctuational mass associated with each of
the $N$ particles in the universe is
$$\frac{\sqrt{N} m}{N} = \frac{m}{\sqrt{N}} \sim 10^{-65}gms,$$ that
is once again the smallest observable mass or ground state mass in the
universe.\\
\section{Further Considerations}
1. We will now provide yet another rationale for (3). Let us
   start with equations encountered earlier, 
$$R = \sqrt{N} l$$
$$\frac{Gm^2}{e^2} = \frac{1}{\sqrt{N}} \sim 10^{-40}$$ or the
Weinberg formula
$$m = \left(\frac{H \hbar^2}{Gc}\right)^{\frac{1}{3}}$$ where $N \sim
10^{80}$ is the number of elementary particles, typically pions, in
the universe. On the other hand the ratio of the
electromagnetic and Gravitational coupling constants, is deducible
from (3). The very mysterious feature of the 'Weinberg'' formula  was stressed by Weinberg himself  ``...it should be noted that the particular combination of
$\hbar , H, G$, and $c$ appearing (in the formula) is very much closer
to a typical elementary particle mass than other random combinations
of these quantities.... \\ In contrast, (the formula) relates a single
cosmological parameter, $H$, to the fundamental constants $\hbar , G,
c$ and $m$, and is so far unexplained...''\\ Relations like (9) and
(8) inspired the Dirac Large Number Cosmology. All these relations are
to be taken in the order of magnitude sense.\\ We will now take a
different route and provide an alternative theoretical rationale for
equations (71), (8) and (9), and in the process light will be shed on
the new cosmological model and the nature of gravitation.\\ Following
Sivaram \cite{sivaram2} we consider the gravitational self energy of
the pion. This is given by
$$\frac{Gm^2}{l} = Gm^2/(\hbar/mc)$$ If this energy were to have a
life time of the order of the age of the universe, $T$, then we have
by the Uncertainty relation
\begin{equation}
\left(\frac{Gm^3c}{\hbar}\right) (T) \approx \hbar\label{ey4}
\end{equation}
As $T = \frac{1}{H}$, this immediately gives us the Weinberg formula
(7). It must be observed again that (\ref{ey4}) gives a time dependent
gravitational constant $G$.\\ We could also derive (7) by using a
relation given by Landsberg \cite{land}. We use the fact that the
mass of a particle is given by
\begin{equation}
m(b) \sim \left(\frac{\hbar^3 H}{G^2}\right)^{1/5}
\left(\frac{c^5}{\hbar H^2 G}\right)^{b/15}\label{ey5}
\end{equation}
where $b$ is an unidentified constant. Whence we have
$$m(b) \sim G^{-3/5} G^{-3b/15} = G^{-(b+1)/5}$$ The mass that would
be time independent, if $G$ were time dependent would be given by the
value
$$b = -1$$ With this value of $b$ (\ref{ey5}) gives back (7).\\ Let us
now proceed along a different track. We rewrite (\ref{ey4}) as
\begin{equation}
G = \frac{\hbar^2}{m^3c} \cdot \frac{1}{T}\label{ey6}
\end{equation}
If we use the fact that $R = cT$, then (\ref{ey6}) can be written as
\begin{equation}
G = \frac{\hbar^2}{m^3R}\label{ey7}
\end{equation}
Let us now use the well known relation encountered earlier
\cite{ruffini}
\begin{equation}
R = \frac{GM}{c^2},\label{ey8}
\end{equation}
There are several derivations of (\ref{ey8}). For example in a
uniformly expanding Friedman universe, we have
$$\dot {R}^2 = \frac{8\pi G \rho R^2}{3}$$ If we substitute the value
$\dot {R} = c$ at the radius of the universe, then we recover
(\ref{ey8}). If we use (\ref{ey8}) in (\ref{ey7}) we will get
\begin{equation}
G^2 = \frac{\hbar^2 c^2}{m^3 M}\label{ey9}
\end{equation}
Let $M/m = N$ be called the number of elementary particles in the
universe. Infact this is just (1). Then (\ref{ey9}) can be written as
(10),
$$G = \frac{\hbar c}{m^2 \sqrt{N}}$$ which can also be written as (8)
$$Gm^2/e^2 \sim \frac{1}{\sqrt{N}}$$ Whence we get (9)
$$\sqrt{N} l = R$$ We now remark that (\ref{ey6}) shows an inverse
dependence on time of the gravitation constant, while (10) shows an
inverse dependence on $\sqrt{N}$. Equating the two, we get back,
$$T = \sqrt{N} \tau$$ the relation (4) which we have encountered
several times. If we now take the time derivative of (10) and use (4),
we get (3)
$$\dot {N} = \frac{\sqrt{N}}{\tau}$$ This equation is the same as (78)
or (3). To put it briefly in a phase transition from the Quantum
Vaccuum $\sqrt{N}$ particles appear within the Compton time $\tau$. In
terms of our unidirectional concept of time, we could say that
particles appear and disappear, but the nett result is the appearance
of $\sqrt{N}$ particles.\\ We now make a few remarks. Firstly it is
interesting to note that $\sqrt{N}m$ will be the mass added to the
universe. Let us now apply the well known Beckenstein formula for the
life time of a mass $M$ viz., \cite{ruffini},
$$t \approx G^2M^3/\hbar c^4$$ to the above mass. The life time as can
be easily verified turns out to be exactly the age of the universe!\\
A final remark. To appreciate the role of fluctuations in the
otherwise mysterious Large Number relations, let us follow Hayakawa
 and consider the excess of electric energy due to the
fluctuation $\sim \sqrt{N}$ of the elementary particles in the
universe and equate it to the inertial energy of an elementary
particle. We get
$$\frac{\sqrt{N}e^2}{R} = mc^2$$ This gives us back (8) if we use
(\ref{ey8}). If we use (9) on the other hand, we get
$$e^2/mc^2 = l,$$ another well known relation from micro physics.\\
2. We note that as is well known, a background ZPF of the kind we have
been considering can explain the Quantum Mechanical spin half as also
the anomalous $g = 2$ factor for an otherwise purely classical
electron \cite{sachi,boyer,taylor}. The key point here is
(Cf.ref.\cite{sachi}) that the classical angular momentum $\vec r
\times m \vec v$ does not satisfy the Quantum Mechanical commutation
rule for the angular momentum $\vec J$. However when we introduce the
background Zero Point Field, the momentum now becomes
\begin{equation}
\vec J = \vec r \times m=\vec v + (e/2c) \vec r \times (\vec B \times
\vec r) + (e/c) \vec r \times \vec A^0 ,\label{ez5}
\end{equation}
where $\vec A^0$ is the vector potential associated with the ZPF and
$\vec B$ is an external magnetic field introduced merely for
convenience, and which can be made vanishingly small.\\ It can be
shown that $\vec J$ in (\ref{ez5}) satisfies the Quantum Mechanical
commutation relation for $\vec J \times \vec J$. At the same time we
can deduce from (\ref{ez5})
\begin{equation}
\langle J_z \rangle = - \frac{1}{2} \hbar
\omega_0/|\omega_0|\label{ez6}
\end{equation}
Relation (\ref{ez6}) gives the correct Quantum Mechanical results
referred to above.\\ From (\ref{ez5}) we can also deduce that
\begin{equation}
l = \langle r^2 \rangle^{\frac{1}{2}} =
\left(\frac{\hbar}{mc}\right)\label{ez7}
\end{equation}
After several fruitless decades of attempts to provide a unified description of Gravitation and Electromagnetism, it has been realized that differentiable space time has to be abandoned in favour of one with a minimum scale. It is now generally accepted that the Planck scale defines the minimum scale for the universe \cite{rr2,rr1,garay,crane,rr25,r2}. In these schemes there is a maximal mass, the Planck mass $m_P \sim 10^{-5}gms$ which is defined by
\begin{equation}
m_P = \left(\frac{\hbar c}{G}\right)^{1/2} \approx 10^{-5}gms\label{ev1}
\end{equation}
Using the value for $m_P$ we can define the Planck length $l_P \sim 10^{-33}cms$ and the Planck time $t_P \sim 10^{-42}secs$, which are the Compton lengths and times for the mass in (\ref{ev1}). It may be mentioned that these values were postulated by Planck himself. Today the values for the minimum scale as given in (\ref{ev1}) are taken for granted. We first provide a rationale for the numerical value of the Planck scale.\\
\section{The Planck Scale}
Our starting point is the model for the underpinning at the Planck scale for the universe. This is a collection of $N$ Planck scale oscillators (Cf.refs.\cite{rr25,r2,uof,gip,ng} for details). We do not need to specfify $N$. We have in this case the following well known relations
$$R = \sqrt{N}l, Kl^2 = kT,$$
\begin{equation}
\omega^2_{max} = \frac{K}{m} = \frac{kT}{ml^2}\label{ev2}
\end{equation}
In (\ref{ev2}), $R$ is of the order of the diameter of the universe, $K$ is the analogue of the spring constant, $T$ is the effective temperature while $l$ is the analogue of the Planck length, $m$ the analogue of the Planck mass and $\omega_{max}$ is the frequency-the reason for the subscript $max$ will be seen below. We do not yet give $l$ and $m$ their usual values as given in (\ref{ev1}) for example, but rather try to deduce these values.\\
We now use the well known result that the individual minimal oscillator particles are black holes or mini universes as shown by Rosen \cite{rr5}. So using the well known Beckenstein temperature formula for these primordial black holes \cite{ruffini}, that is
$$kT = \frac{\hbar c^3}{8\pi Gm}$$
in (\ref{ev2}) we get,
\begin{equation}
Gm^2 \sim \hbar c\label{ev3}
\end{equation}
which is another form of (\ref{ev1}). We can easily verify that (\ref{ev3}) leads to the value $m \sim 10^{-5}gms$. In deducing (\ref{ev3}) we have used the typical expressions for the frequency as the inverse of the time - the Compton time in this case and similarly the expression for the Compton length. However it must be reiterated that no specific values for $l$ or $m$ were considered in the deduction of (\ref{ev3}).\\
We now make two interesting comments. Cercignani and co-workers have shown \cite{cer1,cer2} that when the gravitational energy becomes of the order of the electromagnetic energy in the case of the Zero Point oscillators, that is
\begin{equation}
\frac{G\hbar^2 \omega^3}{c^5} \sim \hbar \omega\label{ev4}
\end{equation}
then this defines a threshold frequency $\omega_{max}$ above in which the oscillations become chaotic.\\
Secondly from the parallel but unrelated theory of phonons \cite{huang,rief}, which are also bosonic oscillators, we deduce a maximal frequency given by
\begin{equation}
\omega^2_{max} = \frac{c^2}{l^2}\label{ev5}
\end{equation}
In (\ref{ev5}) $c$ is, in the particular case of phonons, the velocity of propagation, that is the velocity of sound, whereas in our case this velocity is that of light. Frequencies greater than $\omega_{max}$ in (\ref{ev5}) are meaningless.  We can easily verify that (\ref{ev4}) and (\ref{ev5}) give back (\ref{ev3}).\\
Finally we can see from (\ref{ev2}) that, given the value of $l_P$ and using the value of the radius of the universe, viz., $R \sim 10^{27}$, we can deduce that, 
\begin{equation}
N' \sim 10^{120}\label{ev6}
\end{equation}
In a sense the relation (\ref{ev3}) can be interpreted in a slightly different vein as representing the scale at which all energy- gravitational and electromagnetic becomes one.
\section{The Gauge Hierarchy Problem}
A long standing puzzle has been the so called Gauge Hierarchy Problem. This deals with the fact that the Planck mass is some $10^{20}$ times the mass of an elementary particle, for example Gauge Bosons or Protons or Electrons (in the large number sense). Why is there such a huge gap?\\
We now recall that as already shown we have \cite{rr8,rr9,fpl,uof}
\begin{equation}
G = \frac{\hbar c}{m^2\sqrt{N}}\label{eva3}
\end{equation}
In (\ref{eva3}) $N \sim 10^{80}$ is the well known number of elementary particles in the universe, which features in the Weyl-Eddington relations as also the Dirac Cosmology.\\
What is interesting about (\ref{eva3}) is that it shows gravitation as a distributional effect over all the $N$ particles in the universe \cite{fpl,uof}.\\
Let us rewrite (\ref{ev3}) in the form
\begin{equation}
G \approx \frac{\hbar c}{m^2_P}\label{eva4}
\end{equation}
remembering that the Planck length is also the Compton length of the Planck mass. (Interestingly an equation like (\ref{e3}) or (\ref{ea4}) also follows from Sakharov's treatment of gravitation \cite{sakharov}.) A division of (\ref{ea3}) and (\ref{ea4}) yields
\begin{equation}
m^2_P = \sqrt{N} m^2\label{eva5}
\end{equation}
Equation (\ref{eva5}) immediately gives the ratio $\sim 10^{20}$ between the Planck mass and the mass of an elementary particle.\\
Comparing (\ref{eva3}) and (\ref{eva4}) we can see that the latter is the analogue of the former in the case $N \sim 1$. So while the Planck mass in the spirit of Rosen's isolated universe and the Schwarzchild black hole uses the gravitational interaction in isolation, as seen from (\ref{eva3}), elementary particles are involved in the gravitational interaction with all the remaining particles in the universe.\\
Finally rememebring that $Gm_P^2 \sim e^2$, as can also be seen from (\ref{eva4}), we get from (\ref{eva3})
\begin{equation}
\frac{e^2}{Gm^2} \sim \frac{1}{\sqrt{\bar{N}}}\label{eva6}
\end{equation}
Equation (\ref{eva6}) is the otherwise empirically well known electromagnetism-gravitation coupling constant ratio, but here it is deduced from the theory.\\
It may be remarked that one could attempt an explanation of (\ref{eva5}) from the point of view of SuperSymmetry or Brane theory, but these latter have as yet no experimental validation \cite{gor}. On the other hand, the crucial equation (\ref{eva3}) was actually part of a predicted dark energy driven accelerating universe with a small cosmological constant (besides a deduction of hitherto empirical Large Number relations) (Cf. for example ref.\cite{rr8})-- all this got dramatic observational confirmation through the works of Perlmutter, Kirshner and others, shortly thereafter.

\end{document}